\documentclass{PoS}

\title{Effect of pairing on nuclear dynamic}

\ShortTitle{Effect of pairing on nuclear dynamics}

\author{\speaker{Guillaume Scamps}\\
        GANIL, CEA/DSM and CNRS/IN2P3, Bo\^ite Postale 55027, 14076 Caen Cedex, France\\
        E-mail: \email{scamps@ganil.fr}}

\author{Denis Lacroix\\
        GANIL, CEA/DSM and CNRS/IN2P3, Bo\^ite Postale 55027, 14076 Caen Cedex, France\\
        E-mail: \email{lacroix@ganil.fr}}

\abstract{
The effect of pairing on small and large amplitude dynamics is discussed. Pairing correlation is treated in a fully microscopic transport theory 
using a simplified BCS version of the Time-Dependent Hartree-Fock Bogolyubov (TDHFB) theory.  The approach is applied to study the 
Giant Dipole and Giant Quadrupole Resonance in spherical and/or deformed magnesium isotopes showing very good agreement with QRPA theory.  The present framework also provides a consistent approach to describe nuclear reactions around the Coulomb barrier. As an illustration, the influence of pairing correlations on multi-nucleon transfer is studied for reactions at sub-barrier energy. }

\FullConference{51st International Winter Meeting on Nuclear Physics\\
		 21-25 January 2013\\
		 Bormio (Italy)}

\begin{document}

\section{Introduction}

Pairing correlation is known to affect significantly ground state and excited state spectroscopy of nuclei. On the theory side, static 
properties of nuclei are commonly treated using the  
Hartree-Fock-Bogoliubov (HFB) approach within the Energy Density Functional (EDF) theory. The Quasi-Particle RPA (QRPA) 
extend the HFB method and provides a consistent approach allowing to describe both low lying excited states and giant resonances.
Due to the underlying small amplitude approximation, it cannot treat large amplitude collective motion.  Large effort is currently made 
to develop the Time-Dependent HFB (TDHFB) from which the QRPA can be derived. At present, very few applications 
of TDHFB exist so far \cite{Has07,Ave08,Ste11}  and most of them have been made 
to study giant resonances where the QRPA already provides a competitive approach. The TDHFB theory is very demanding numerically 
and increases the computer time by a factor from 100 to 1000 compared to the corresponding theory without pairing. This might explain 
why such approach has been up to now mainly restricted to the description of a single nucleus presenting small oscillations around equilibrium. In a series of work\cite{Eba10,Sca12,Sca13}, it has been shown that the BCS version of TDHFB can be a good compromise between the required numerical effort and the description of pairing beyond the independent particle picture both for giant resonances 
and nuclear collisions.  Some illustrations of recent achievements are shown below.
  
\section{Time-dependent dynamic with pairing}

In the present approach, the many-body trial wave-function is written as a quasi-particle state 
in the BCS form 
\begin{equation}
|\Psi\rangle = \prod_{k>0} \left( u_k(t) + v_k(t) a^{\dagger}_k(t) a^{\dagger}_{\overline k}(t) \right) | - \rangle, \label{eq:tdstate}
\end{equation}
where $u_k(t)$ and $v_k(t)$ are the components of the special Bogoliubov transformation linking the 
quasi-particle creation/annihilation to the particle creation/annihilation of the canonical states, denoted by $\{a^{\dagger}_k(t), 
a^{\dagger}_{\bar k}(t)\}$. This operators are associated to components $\varphi_k({\bf r}, \sigma) $ in {\bf r}-space and spin space with
$a^{\dagger}_k = \sum_{\sigma} \int d{\bf r} \varphi_k({\bf r}, \sigma) \Psi^{\dagger}_{\sigma}({\bf r})$. Starting from the TDHFB equation 
and neglecting the off-diagonal part of the pairing field, leads to the TDHF+BCS approximation.  The equations of motion are then conveniently written in terms of the occupation numbers $n_k (t)= v^2_k(t)$ of single-particle states 
and  anomalous density components $\kappa_k(t) = u_k^*(t) v_k(t)$:
\begin{equation}
i\hbar\frac{d n_k}{dt} = \Delta^*_k \kappa_k - \Delta_k \kappa^*_k, \quad
i\hbar\frac{d \kappa_k}{dt} =  \kappa_i(\epsilon_k-\epsilon_{\overline{k}}) + \Delta_k(2n_k-1)
\end{equation}
where $\Delta_k$ is the pairing field. The advantage of TDHF+BCS compared to TDHFB is that canonical states 
evolves according to a TDHF like equation of motion:
\begin{equation}
i\hbar \frac{d}{dt}|\varphi_k\rangle = (h[\rho] - \epsilon_k(t)) |\varphi_k\rangle 
\end{equation}
where $h[\rho]$ is the self-consistent mean-field while 
$\epsilon_k$ is a factor that is conveniently chosen as $\epsilon_k = \langle \varphi_k |h[\rho]| \varphi_k \rangle $.
Properties of TDHF+BCS, also called Canonical basis TDHFB (CbTDHFB) have been extensively discussed in ref. \cite{Eba10,Sca12}.  
Such simplified approach has clear numerical advantages but might also lead to inconsistencies related to continuity equations \cite{Sca12}. For this reason, it might be also interesting to consider an extra simplification where occupation numbers and correlations 
are frozen in time (the Frozen Occupation Approximation (FOA)). In the following, three theories are compared: the TDHF approach, the 
TDHF+BCS and the TDHF+BCS in the FOA limit.  The transport equations are solved on a 3D mesh using a Skyrme functional in the mean-field and a contact interaction in the pairing channel (for more details see ref. \cite{Sca12}). 

\section{Effect of pairing on collective vibrations in nuclei}

\begin{figure}[!ht] 
	\centering\includegraphics[width=0.6\linewidth]{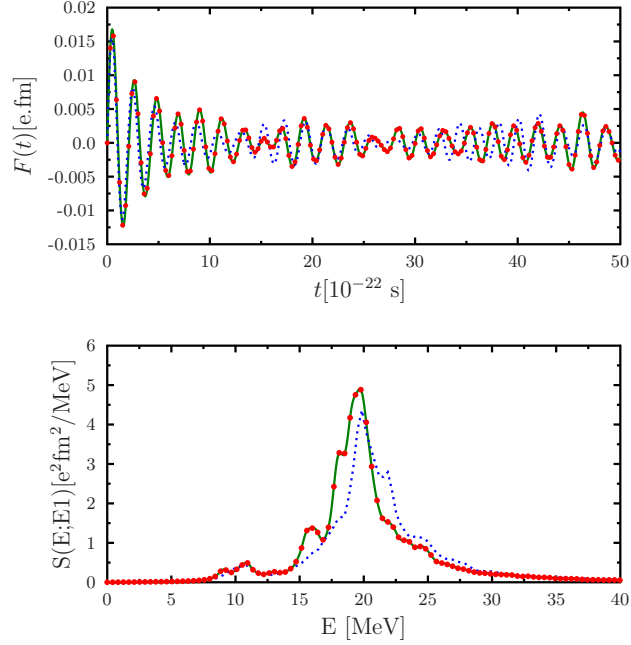} 
	\caption{Top: Illustration of the dipole moment evolution as a function of time in $^{28}$Mg, after an initial IV-GDR boost. 
	Bottom: Strength function for $^{28}$Mg. Both panels present three different theories, TDHF (dashed line), FOA (solid line) and TDHF+BCS (dots). 
	The calculations have been performed using the Skyrme Skm* parameters and a mixed pairing interaction.}	
	\label{fig:FOA_HF_G2_Yb172_dip} 
\end{figure}

To study the collective response of nuclei, we consider an initial state $| \Psi(t_0) \rangle = e^{-i\eta \hat F /\hbar} |\Psi_0 \rangle$ 
where $|\Psi_0 \rangle$ is the quasi-particle ground state that is stationary solution of the BCS equation. The coefficient $\eta$ should be small 
enough to insure the small amplitude hypothesis. The operator $\hat F$ is chosen such that specific collective modes are excited. 
In the following, we will consider the isovector Giant Dipole resonance (IV-GDR), and the isoscalar Giant Quadrupole Resonance (IS-GQR). Explicit form of the associated operators can be found in ref.  \cite{Eba10}. The nucleus response is obtained by solving the TDHF+BCS leading to time oscillations of $F(t)=\langle \Psi(t)|{\hat F} | \Psi(t) \rangle$ where $| \Psi(t) \rangle$ is the time-dependent quasi-particle state (\ref{eq:tdstate}). An illustration, of the GDR oscillation is given in Figure \ref{fig:FOA_HF_G2_Yb172_dip}  (top).  The strength function, generally used in RPA and/or QRPA is then linked to the Fourier transform of $F(t)$, denoted by $\tilde{F}(E)$, through 
\begin{equation}
S(E)=\frac1{\eta \pi} {\rm Im} (\tilde{F}(E)).
\end{equation}

Such strength distribution is shown in Fig. \ref{fig:FOA_HF_G2_Yb172_dip} for the TDHF case (no pairing), TDHF+BCS and 
starting from a ground state with pairing but assuming that the pairing degrees of freedom are frozen in time (FOA). Note that all the calculations for giant resonances are done using the Skm$^*$ functional, with a mixed pairing interaction with parameters from \cite{Ber09}.
From the present comparison, one can conclude that the pairing correlations induces a modification of the GDR strength.  In particular, 
a small shift of the main pic energy towards low energy is observed while the width of the giant resonance (life-time) 
is globally unchanged. Such a shift is indeed expected in approaches going beyond the pure independent particle picture.   
It is interesting to note that there is almost no difference between the cases where correlations are propagated in time or are frozen.
This shows that the main effect of pairing on high-lying giant resonance is the initial fragmentation of the occupation numbers around the 
Fermi energy induced by pairing correlations. As we will see below this conclusion does not hold for low lying collective modes.

\begin{figure}[!ht] 
	\centering\includegraphics[width=0.85\linewidth]{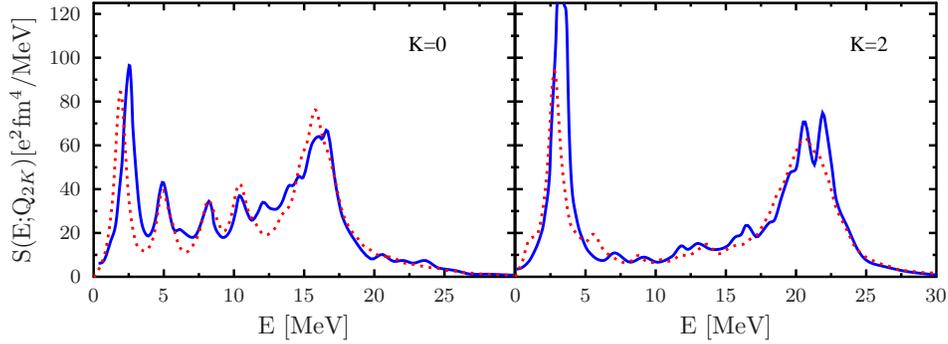} 	
	\caption{Isoscalar quadrupole responses for $^{34}$Mg. Comparison between QRPA calculations \cite{Los10} (solid line) and TDHF+BCS (dashed line). }
	\label{fig:comp_QRPA} 
\end{figure}
As a second example, the TDHF+BCS response to a quadrupole perturbation in $^{34}$Mg is shown in figure \ref{fig:comp_QRPA}. 
Note that, this nucleus is found to be deformed with a deformation parameter $\beta_2=0.33$ in its ground state.  Therefore, the response 
is different depending on the axis direction of the perturbation. The response along the two main axis ($K=0$) and ($K=2$) are respectively
shown on the left and right side of Fig. \ref{fig:comp_QRPA}.  In both cases, the results are compared with the deformed QRPA of 
ref. \cite{Los10}. We see that the TDHF+BCS results, pic position and height, are in rather good agreement with the QRPA 
results although the QRPA is expected to be more general since it corresponds to the full TDHFB. Similar conclusion have been drawn in ref. \cite{Eba10} using a different pairing interaction. 

\begin{figure}[!ht] 
	\centering\includegraphics[width=0.85\linewidth]{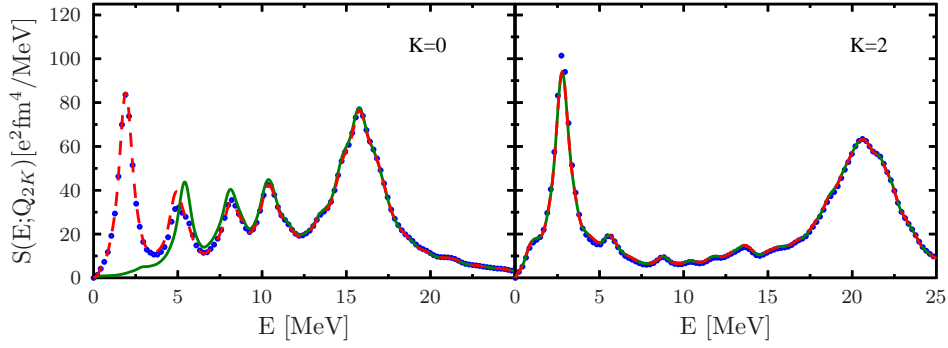} 
		\caption{Isoscalar quadrupole responses for $^{34}$Mg. Comparison between FOA (solid line), the TDHF+BCS (dashed line)
	and the TDHF+BCS dynamics (filled circles) projected on good number of protons and neutrons. 
}
	\label{fig:comp_FOA_TDBCS} 
\end{figure}

It is worth mentioning that the TDHF+BCS results presented in Fig. \ref{fig:comp_QRPA} includes the reorganization of correlations 
in time. The corresponding results where occupation numbers are frozen (FOA) are shown in Fig. \ref{fig:comp_FOA_TDBCS}. Contrary to 
the IV-GDR that was not presenting low lying collective modes, the IS-GQR has a significant fraction of the strength at energy below 5 MeV.
These low energy modes that are present in QRPA are absent in the FOA approximation. Therefore, while the effect of pairing on the 
high-energy collective modes could be explained by the initial fragmentation of the occupation numbers near the Fermi energy, the 
origin of low energy collective modes seems more complex and could only be understood through the propagation of correlation in time.    

The BCS wave function use the powerful technique of symmetry breaking, here the $U(1)$ symmetry associated to particle number conservation, to grasp pairing correlation. Accordingly, the response presented in 
Figs. \ref{fig:FOA_HF_G2_Yb172_dip}-\ref{fig:comp_FOA_TDBCS}  do correspond to a weighted average of the response of nuclei 
with various proton and neutron numbers.  One might worry in that case about the possible pollution of the strength by collective excitation 
of nuclei surrounding the nucleus of interest. To quantify the effect of particle number symmetry breaking, we have developed 
a projection after variation (PAV) method for nuclear dynamics. In this method, the response of the nucleus with proton number $N$ 
and neutron number $Z$ is deduced by estimating the operator $\hat F$ on the projected component of $| \Psi(t) \rangle$, i.e.  
\begin{equation}
F_{N,Z}(t)=\frac{\langle \Psi(t)| \hat{P}(N)\hat{P}(Z) {\hat F} \hat{P}(N) \hat{P}(Z)| \Psi(t) \rangle}{\langle \Psi(t)| \hat{P}(N) \hat{P}(Z)| \Psi(t) \rangle},
\end{equation}
where $\hat{P}(N)$ and $\hat{P}(Z)$ are projectors on good particle number defined through:
\begin{eqnarray}
  \hat{P}(X) = \frac{1}{2\pi}\int_0^{2\pi} e^{i \varphi (\hat X - X)} d\varphi
\end{eqnarray}
with $X=N$ or $Z$.
The quasi-particle states here, follow the TDHF+BCS equation of motion. Similarly to the original case, 
the strength function is obtained by performing the Fourier transform of $F_{N,Z}(t)$. Such a strength is shown in Fig. 
\ref{fig:comp_FOA_TDBCS} by filled circles. Our conclusion is that the effect  of particle non-conservation on the collective response 
seems to be rather weak in general.

\section{Effect of pairing on two-nucleon transfer at sub-barrier energy}

The TDHF+BCS method can be easily implemented to perform nuclear collision using similar technique as for the TDHF approach.
An illustration has been given in ref. \cite{Sca12} where the effect of pairing on single- and multi-nucleon transfer was analyzed.
Recently, there is a renewal of interest regarding the transfer in the sub-barrier regime in particular to understand its possible 
effect as a competing or contributing effect to fusion \cite{Cor11,Sar12}. The understanding of pair transfer in that case is of special interest since it gives some information on how a composite pair of interacting particles might be transferred by tunneling. 
The present approach, by including 
pairing  correlation as well as single-particle quantum dynamics provides a way to describe such process.
    
\begin{figure}[!ht] 
	\centering\includegraphics[width=0.9\linewidth]{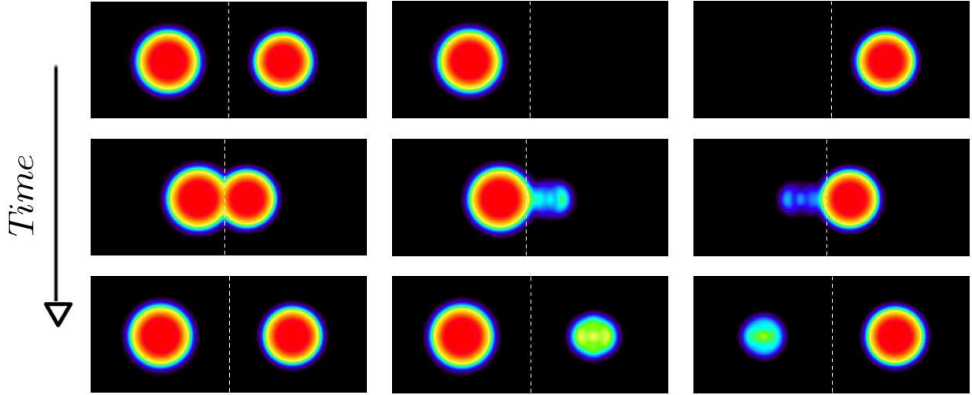} 
	\caption{Neutron density evolution for the central collision $^{48}$Ca (left nucleus)+$^{40}$Ca (right nucleus) at different time: before (top), during (middle) and after the
	contact (bottom). The time evolution of the density of nucleons that are initially in the $^{48}$Ca (resp. $^{40}$Ca) is shown 
	in the middle (resp. right) panel.  On the left panel, we the total neutron density is shown.  }
	\label{fig:pict} 
\end{figure}

In Fig. \ref{fig:pict}, illustration of density profiles evolution during the central collision $^{48}$Ca+$^{40}$Ca are shown for different time of the reaction. The beam energy is 48.6 MeV, that corresponds to a center of mass energy just below the Coulomb barrier $V_B=51.9$ MeV. 
During the collision, the two nuclei approaches each other up to the contact and re-separate. During the contact time, nucleons are 
exchanged as can be seen on middle and right side of Fig. \ref{fig:pict}. In the TDHF framework, multi-nucleon exchange is treated as
an independent particle process while when pairing is included, we do expect in an enhancement of cooperative pair transfer. 
This aspect could be approximately treated with TDHF+BCS. However, due to the break-down of the continuity equation \cite{Sca12}, 
that is especially important when it is necessary to cut a system in two pieces, which is obviously the case for transfer reaction, 
it is more convenient to use the FOA approximation that partially cure this problem.  The results presented below are considering this limit. Therefore, possible effects that are due to internal reorganization of correlations during the time evolution are neglected.  

\begin{figure}[!ht] 
	\centering\includegraphics[width=0.7\linewidth]{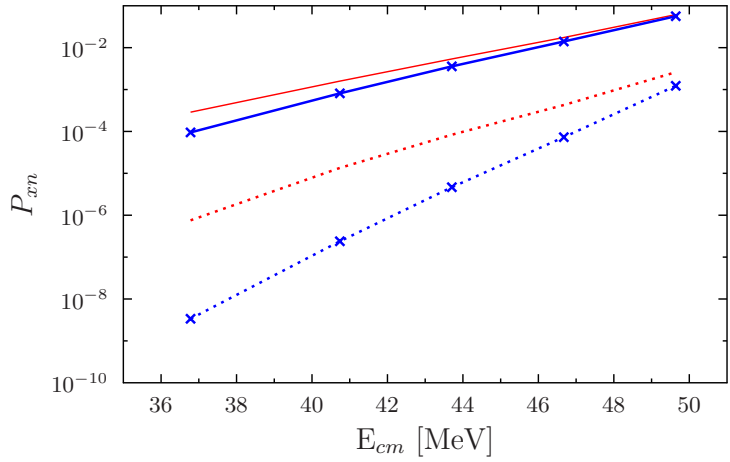} 
	\caption{Illustration of one-neutron (solid lines) and two-neutron (dashed line) transfer probabilities obtained in the reaction 
	$^{48}$Ca+$^{40}$Ca as a function of the center of mass energy. The calculation are done with (lines) and without (lines with crosses)  pairing. }
	\label{fig:comp_Ca40_Ca46} 
\end{figure}

An additional difficulty appears when theories breaking the particle number are used. Indeed, as in the case of giant resonances, for non vanishing pairing, the initial superfluid nucleus should be considered as an average over nuclei with different neutron and proton numbers.
To bypass this problem in the case of transfer reactions, we have proposed in ref. \cite{Sca13} to use a double projection technique. Since we are interested in the final number of nucleons in one of the side of the contact plane, following ref. \cite{Sim10}, 
an operator $N_{\Theta}=\sum_{\sigma}\int d { \bf r} \Psi^{\dagger}({ \bf r},\sigma) \Psi({ \bf r},\sigma) \Theta({ \bf r}) $ counting the 
number of particle in this side is defined. $\Theta({ \bf r}) $ is the Heaviside function equal to $1$ in the considered subspace 
and $0$ otherwise.  Then, one could associate to this operator, a projector %
\begin{equation}
{\hat P}_{\Theta}(N) = \frac1{2\pi} \int_{0}^{2\pi} e^{i \varphi (\hat N_{\Theta} - N)} d\varphi,
\end{equation}
that gives access to the probability to find $N$ particles in the considered space. 
However, contrary to the standard TDHF case, since the initial state does not exactly preserves particle number, one should in
addition define the projector onto the good particle in the total space. Accordingly, the probability to find $N'$ particles in 
one side of the reactions while to total number of particles is assumed to be $N$ is given by 
\begin{equation}
P_{\Theta}(N')=\frac{ \langle \Psi (t) | {\hat P}_{\Theta}(N') {\hat P}(N) | \Psi(t) \rangle }{\langle \Psi (t) | {\hat P}(N) | \Psi (t)\rangle}
\end{equation}
The numerical details of the calculation are presented in Ref \cite{Sca13}. 
An illustration of the enhancement of pair transfer induced by pairing correlations is given in Fig.  \ref{fig:comp_Ca40_Ca46}.
In this figure, the one- and two-particle transfer probabilities obtained using TDHF and TDHF+BCS (FOA) 
are compared for the 
reaction $^{48}$Ca+$^{40}$Ca  for different energies below the Coulomb barrier.  From this figure, two effects of pairing can be establish. First, the one-neutron transfer probability is slightly enhanced, this effect is due to the fragmentation of the occupation number that allows occupation above the fermi energy. A second effect of pairing stems directly from the non-zero initial two-body correlations. Those correlations increase significantly the transfer of pair of neutrons. 

\begin{figure}[!ht] 
	\centering\includegraphics[width=0.6\linewidth]{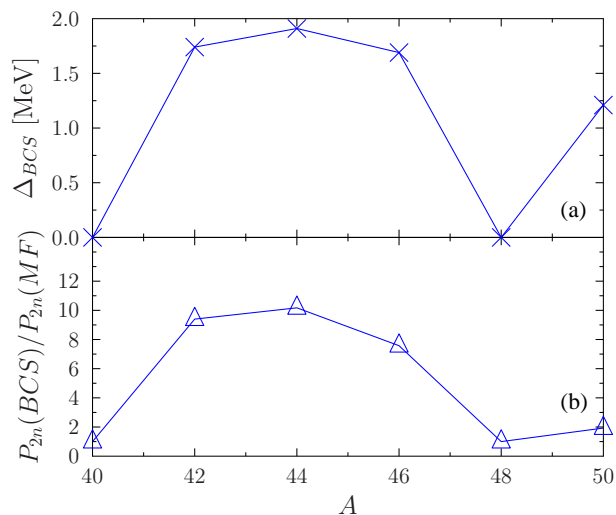} 
	\caption{Panel (a): Mean neutron pairing gap as a function of the mass for the Ca isotopic chain. Panel (b): Enhancement of pair transfer due to pairing as a function of the mass for the reactions $^{X}$Ca+$^{40}$Ca.}
	\label{fig:gap_bell} 
\end{figure}

To further quantify the effect of pairing on the two-particle transfer, a systematic study of reactions $^{X}$Ca+$^{40}$Ca for $X=40$ to $50$ has been made in ref. \cite{Sca13} . An illustration of the connection 
between the pair transfer enhancement and the initial pairing correlation is given in  Fig. \ref{fig:gap_bell}. In this figure, the ratio $P_{2n}(BCS)/P_{2n}(MF)$ for the considered reactions as a function of $X$ and is systematically considered at a center of mass energy at
by 6 MeV smaller than the Coulomb barrier. This ratio represents the enhancement of pair transfer probability due to pairing. 
In the top panel of this figure, the initial pairing gap in the $^{X}$Ca collision partner is also presented. We see that the 
pair transfer enhancement is quantitatively and qualitatively strongly correlated to the initial strength of the pairing correlation 
of the considered nucleus.

\section{Conclusion}

In the present proceedings, several applications of the TDHF+BCS theory to collective motion and giant resonances illustrates 
how pairing can affect the small and large amplitude dynamics in nuclei. It is shown that pairing not only 
significantly affects low lying collective states as anticipated but also slightly shift the energy of giant resonances. 
The TDHF+BCS approach is also applied to study the effect of pairing on two-particle transfer. A strong enhancement 
that significantly depend on the beam energy is observed.

\end{document}